\documentclass[sigconf,authorversion,nonacm]{acmart}

\usepackage{amsmath,amsfonts}
\usepackage{algorithmic}
\usepackage{graphicx}
\usepackage{textcomp}
\usepackage{xcolor}
\usepackage{cleveref}
\usepackage{enumitem}
\usepackage[acronym]{glossaries}
\usepackage{multirow}
\usepackage{xspace}
\usepackage{microtype}
\usepackage{pifont}

\usepackage[free-standing-units,per-mode=repeated-symbol]{siunitx}
\usepackage[allow-number-unit-breaks]{siunitx}
\DeclareSIUnit\flop{FLOP}
\DeclareSIUnit\flops{FLOPS}
\DeclareSIUnit\gate{GE}
\DeclareSIUnit\op{OP}
\DeclareSIUnit\macu{MACU}
\DeclareSIUnit\ops{OPS}
\DeclareSIUnit\core{core}
\DeclareSIUnit\request{request}
\DeclareSIUnit\cycle{cycle}
\DeclareSIUnit\teraops{TOPS}
\DeclareSIUnit\ghz{GHz}
\DeclareSIUnit\mhz{MHz}
\DeclareSIUnit[number-unit-product = ]\percent{\%}

\newcommand\eg{e.g.,\xspace}

\newcommand\terapool[1]{\ensuremath{\text{TeraPool-SDR}_{#1}}}
\newcommand\tp[1]{\ensuremath{\text{TP-SDR}_{#1}}}
\def\checkmark{\text{\ding{51}}}
\makeatletter \newcommand{\AddSpaceIfAnonymous}{\@ifclasswith{acmart}{anonymous}{\vspace{10mm}}{}} \makeatother

\definecolor{MidnightBlue}{HTML}{191970}
\definecolor{Mint}{HTML}{3EB889}
\definecolor{EnglishRed}{HTML}{A4515C}
\definecolor{SelectiveYellow}{HTML}{FFBA08}
\definecolor{CyanProcess}{HTML}{08B2E3}
\definecolor{OliveDrab7}{HTML}{4D4730}
\definecolor{Red}{HTML}{FF0000}
\colorlet{color1}{MidnightBlue}
\colorlet{color2}{Mint}
\colorlet{color3}{EnglishRed}
\colorlet{color4}{SelectiveYellow}
\colorlet{color5}{CyanProcess}
\colorlet{color6}{OliveDrab7}
\colorlet{colorAlert}{Red}

\acmSubmissionID{6003}

\copyrightyear{2024}
\acmYear{2024}
\setcopyright{acmlicensed}\acmConference[GLSVLSI '24]{Great Lakes Symposium on VLSI 2024}{June 12--14, 2024}{Clearwater, FL, USA}
\acmBooktitle{Great Lakes Symposium on VLSI 2024 (GLSVLSI '24), June 12--14, 2024, Clearwater, FL, USA}
\acmDOI{10.1145/3649476.3658735}
\acmISBN{979-8-4007-0605-9/24/06}

\newacronym{pe}{PE}{Processing Element}
\newacronym{ai}{AI}{Artificial Intelligence}
\newacronym{ml}{ML}{Machine Learning}
\newacronym{cpu}{CPU}{Central Processing Unit}
\newacronym{gpgpu}{GP-GPU}{General Purpose Graphics Processing Unit}
\newacronym{gpu}{GPU}{Graphics Processing Unit}
\newacronym{sm}{SM}{Streaming Multiprocessor}
\newacronym{asic}{ASIC}{Application Specific Integrated Circuit}

\newacronym{5g}{5G}{5th-Generation}
\newacronym{6g}{6G}{6th-Generation}
\newacronym{pusch}{PUSCH}{Physical Uplink Shared Channel}
\newacronym{sdr}{SDR}{Software Defined Radio}
\newacronym{phy}{PHY}{Physical Layer}
\newacronym{ofdma}{OFDMA}{Orthogonal Frequency Division Multiple Access}
\newacronym{bf}{BF}{Beam Forming}
\newacronym{mimo}{MIMO}{Multiple-Input, Multiple-Output}
\newacronym{oran}{O-RAN}{Open Radio Access Networks}
\newacronym{ran}{RAN}{Radio Access Networks}
\newacronym{cots}{COTS}{Commercial-Off-The-Shelf}

\newacronym{axpy}{AXPY}{A Times X Plus Y}
\newacronym{dotp}{DOTP}{Dot Product}
\newacronym{matmul}{MatMul}{Matrix Multiplication}
\newacronym{fft}{FFT}{Fast Fourier Transform}
\newacronym{che}{CHE}{Channel Estimation}
\newacronym{sysinv}{SysInv}{Linear System Inversion}
\newacronym{choldec}{CholDec}{Cholesky Decomposition}
\newacronym{ofdm}{OFDM}{Orthogonal Frequency Division Multiplexing}
\newacronym{tti}{TTI}{Transition Time Interval}

\newacronym{ppa}{PPA}{Power, Performance and Area}
\newacronym{soc}{SoC}{System-on-Chip}
\newacronym{spm}{SPM}{Scratchpad Memory}
\newacronym{numa}{NUMA}{Non-Uniform Memory Access}
\newacronym{isa}{ISA}{Instruction Set Architecture}
\newacronym{fpu}{FPU}{Floating Point Unit}
\newacronym{ipu}{IPU}{Integer Processing Unit}
\newacronym{dsp}{DSP}{Digital Signal Processing}
\newacronym{simd}{SIMD}{Single Instruction Multiple Data}
\newacronym{eda}{EDA}{Electronic Design Automation}
\newacronym{tcdm}{TCDM}{Tightly Coupled Data Memory}
\newacronym{ge}{GE}{Gate Equivalent}
\newacronym{fo4}{FO4}{Fan-Out-of-4}
\newacronym{IDol}{I\$}{Instruction Cache}
\newacronym{sram}{SRAM}{Static Random-Access Memory}
\newacronym{dma}{DMA}{Direct Memory Access}
\newacronym{add}{add}{Add}
\newacronym{mul}{mul}{Multiply}
\newacronym{mac}{MAC}{Multiply Accumulate}
\newacronym{pmac}{p.mac}{Post-increment Multiply-accumulate}

\newacronym{ipc}{IPC}{instructions-per-cycle}
\newacronym{wfi}{WFI}{wait-for-interrupt}
\newacronym{raw}{RAW}{read-after-write}
\newacronym{lsu}{LSU}{Load Store Unit}
\newacronym{ins}{INS}{instruction}
\newacronym{axi}{AXI}{Advanced eXtensible Interface}
\newacronym{beol}{BEOL}{Back-End-of-Line}
\newacronym{pnr}{PnR}{Place and Route}
\newacronym{fc}{FC}{Fully-Connected}
\newacronym{soa}{SoA}{state-of-the-art}

\begin{document}

\title[TeraPool-SDR: A Shared-L1 Cluster for Next-Generation Software-Defined Radios]{TeraPool-SDR: An 1.89TOPS 1024 RV-Cores 4MiB Shared-L1 Cluster for Next-Generation Open-Source Software-Defined Radios}






\author{Yichao Zhang}
\email{yiczhang@iis.ee.ethz.ch}
\orcid{0009-0008-7508-599X}
\affiliation{%
  \institution{Integrated Systems Laboratory}
  \institution{ETH Z\"urich}
  \city{Z\"urich}
  \country{Switzerland}
}

\author{Marco Bertuletti}
\email{mbertuletti@iis.ee.ethz.ch}
\orcid{0000-0001-7576-0803}
\affiliation{%
  \institution{Integrated Systems Laboratory}
  \institution{ETH Z\"urich}
  \city{Z\"urich}
  \country{Switzerland}
}

\author{Samuel Riedel}
\email{sriedel@iis.ee.ethz.ch}
\orcid{0000-0002-5772-6377}
\affiliation{%
  \institution{Integrated Systems Laboratory}
  \institution{ETH Z\"urich}
  \city{Z\"urich}
  \country{Switzerland}
}

\author{Matheus Cavalcante}
\email{matheusd@iis.ee.ethz.ch}
\orcid{0000-0001-9199-1708}
\affiliation{%
  \institution{Integrated Systems Laboratory}
  \institution{ETH Z\"urich}
  \city{Z\"urich}
  \country{Switzerland}
}

\author{Alessandro Vanelli-Coralli}
\email{avanelli@iis.ee.ethz.ch}
\orcid{0000-0002-4475-5718}
\affiliation{%
  \institution{Integrated Systems Laboratory}
  \institution{ETH Z\"urich}
  \city{Z\"urich}
  \country{Switzerland}
}
\affiliation{%
  \institution{Universit\`a di Bologna}
  \city{Bologna}
  \country{Italy}
}

\author{Luca Benini}
\email{lbenini@iis.ee.ethz.ch}
\orcid{0000-0001-8068-3806}
\affiliation{%
  \institution{Integrated Systems Laboratory}
  \institution{ETH Z\"urich}
  \city{Z\"urich}
  \country{Switzerland}
}
\affiliation{%
  \institution{Universit\`a di Bologna}
  \city{Bologna}
  \country{Italy}
}

\renewcommand{\shortauthors}{Zhang Y. and Bertuletti M., et al.}

\begin{abstract}
\gls{ran} workloads are rapidly scaling up in data processing intensity and throughput as the 5G (and beyond) standards grow in number of antennas and sub-carriers. Offering flexible \glspl{pe}, efficient memory access, and a productive parallel programming model, many-core clusters are a well-matched architecture for next-generation software-defined \glspl{ran}, but staggering performance requirements demand a high number of \glspl{pe} coupled with extreme \gls{ppa} efficiency. We present the architecture, design, and full physical implementation of Terapool-SDR, a cluster for \gls{sdr} with 1024 latency-tolerant, compact RV32 \glspl{pe}, sharing a global view of a \qty{4}{\mebi\byte}, \num{4096}-banked, L1 memory. We report various feasible configurations of TeraPool-SDR featuring an ultra-high bandwidth \gls{pe}-to-L1-memory interconnect, clocked at \qty{730}{\mega\hertz}, \qty{880}{\mega\hertz}, and \qty{924}{\mega\hertz} (TT/\SI{0.80}{\volt}/\SI{25}{\celsius}) in \SI{12}{\nm} FinFET technology. The TeraPool-SDR cluster achieves high energy efficiency on all \gls{sdr} key kernels for 5G \glspl{ran}: Fast Fourier Transform (\qty{93}{\giga\ops\per\watt}), Matrix-Multiplication (\qty{125}{\giga\ops\per\watt}), Channel Estimation (\qty{96}{\giga\ops\per\watt}), and Linear System Inversion (\qty{61}{\giga\ops\per\watt}). For all the kernels, it consumes less than \qty{10}{\watt}, in compliance with industry standards.
\AddSpaceIfAnonymous
\end{abstract}



\maketitle

\glsresetall{}
\section{Introduction}
Implementing the \gls{5g} \gls{ran} standards and beyond, requires flexible high-performance computing systems that can sustain large computational workloads and memory footprints in a tight power envelope. To guarantee time-to-market and performance, industry is moving from architectures where network functions are offloaded to specialized accelerators to an open and disaggregated \gls{sdr} paradigm~\cite{Mitola_SDR_IEEE_1995}, building processing pipelines as a chain of programmable components. Off-the-shelf products include programmable~\cite{edgeq_EdgeQ, qualcomm_QRUX100, marvell_octeon10} or reconfigurable~\cite{xilinx_ZYNQ_RFSOC} hardware with their in-house software libraries. Their designs target \SI{10}{Gbps} uplink bandwidth at less than \SI{50}{\watt} power consumption~\cite{edgeq_EdgeQ, qualcomm_QRUX100}, on \gls{fft} for \gls{ofdm}, \gls{matmul} for \gls{bf}, \gls{che}, and \gls{sysinv} for \gls{mimo} detection: the big heterogeneous workloads of the 7.X \gls{oran} functional splits \cite{Larsen_5Gsplits_IEEE_2019}. However, closed proprietary does not allow open research, community-based hardware-software co-design, a propulsive force for \glspl{sdr}.

Focusing on programmable solutions, the many-core cluster offers energy-efficient parallelism. A successful architectural pattern is shared-L1 cluster that eliminates the hardware and software overheads incurred by replicating a private-L1-base many-core cluster for performance purposes~\cite{ET_SoC_1, Ramon_2021} (synchronization, inter-cluster data allocation-splitting, and workload distribution among clusters). Increasing the shared memory many-core cluster scale is also highly desirable to exploit the embarrassingly parallel features of \gls{sdr} processing. Focusing on a concrete example, the parallelization of \gls{5g} \gls{pusch} receiver processing~\cite{PUSCH_2023_DATE} exhibits strong data dependencies: for instance, different \gls{fft} streams of the \gls{ofdm} stage need to be merged and multiplied by a matrix of coefficients in the \gls{bf} stage. If a single cluster shared-L1 is not large enough, \glspl{matmul} can be tiled, but the \gls{ofdm} data must first be merged through the upper levels of the memory hierarchy. \Cref{tab:pusch} resumes this data transfer overhead for each stream, comparing four loosely-coupled \SI{1}{\mebi\byte} clusters and a single shared-L1 \SI{4}{\mebi\byte} cluster, working on \num{64} \gls{ofdm}-antennas of \num{3276}-subcarriers and a \numproduct{32x64x3276} \gls{bf} \gls{matmul}. In the first case, the \SI{32}{b} words \num{64} \gls{ofdm}-antennas are divided into groups of \num{16}, the output is transferred to L2 and the \gls{matmul} is tiled over rows for each processing streams. In the second, the \num{64} antennas are all processed in L1 and the \gls{matmul} stage only requires the transfer of coefficients.
\begin{table}[htb]
  \caption{The data transfer for 5G PUSCH in different clusters.}
  \setlength{\tabcolsep}{2pt}
  \resizebox{\linewidth}{!}{%
    \begin{tabular}[h]{rcc}
      \toprule & \multicolumn{1}{c}{\SI{1}{\mebi\byte} Clusters} & \multicolumn{1}{c}{\SI{4}{\mebi\byte} Cluster} \\
      \cmidrule(l){2-2} \cmidrule(l){3-3}    
      {\#Clusters$\times$Cores}                          & {4$\times$256}            & {1$\times$1024}  \\ 
      {(OFDM-antennas$\times$subcarriers)/Cluster}       & \numproduct{16x3276}      & \numproduct{64x3276}   \\
      {BF(MatMul) Size/Cluster}                          & \numproduct{32x64x819}    & \numproduct{32x64x3276}\\
      {Max. L1-occupation (\si{\kibi\byte})}             & {520}          & {2055}      \\ \midrule
      {Transfer-out OFDM-antennas (\si{\kibi\byte})}     & {205}          & {-}         \\
      {Transfer-in BF-inputs    (\si{\kibi\byte})}       & {213}          & {8}         \\
      {Total Transfer Overhead (\si{\kibi\byte})}        & {418}          & {8}         \\ \midrule
    \end{tabular}}
  \label{tab:pusch}
\end{table}

The existing many-core cluster only up to tens complex processors with shared L1 space: Kalray's MPPA-256 features a 16-cores shared \SI{2}{\mebi\byte} cluster-based architecture, Esperanto ET-SoC-1’s \SI{4}{\mebi\byte} L1 are shared by \num{32} vector cores and Ramon RC64 has \num{64} cores with \SI{4}{\mebi\byte} shared memory~\cite{kalary_256, ET_SoC_1, Ramon_2021}, which are not scalable to the parallel processing of \gls{5g} \gls{oran} large problems. 
The \gls{ofdm} and \gls{bf} are indeed very compute-intensive steps. With $N_{SC}$ sub-carriers, $N_R$ antennas, $N_B$ beams, the computational complexity of the kernels is $O(N_R \times N_{SC}log(N_{SC}) + N_R \times N_{SC} \times N_B)$ real \gls{mac} operations. In \gls{pusch}, up to \num{14} \gls{fft} \& \gls{matmul} streams are executed per \num{1}ms \gls{tti}, which in a typical use-case ($N_{SC}=3276$, $N_B=32$), takes $\sim${\SI{0.8}{\teraops}} with \num{64} receivers (\gls{5g}) and $\sim${\SI{1.8}{\teraops}} with \num{128} (6G) receivers. This requires a scalable cluster with thousands of programmable agile cores, running at near GHz frequency.

This work introduces TeraPool-SDR, a peak \SI{1.89}{\tera\ops} cluster, featuring \num{1024} RISC-V cores with hierarchical low-latency interconnections to a fully shared \SI{4}{\mebi\byte}, \num{4096}-banked L1 \gls{spm}.
The contributions are:
\begin{itemize}[leftmargin=*]{
    \item A physical-aware architecture design for \gls{oran} workloads, based on a three-level physical implementation hierarchy: the Tile, the SubGroup, and the Group;
    \item A detail latency-throughput evaluation of on-chip \gls{numa} latency interconnection;
    \item A complete physical implementation and \gls{ppa} analysis of each design configuration, using the cutting-edge \SI{12}{\nano\meter} FinFET technology node;
    \item A comprehensive performance and energy efficiency evaluation on key \gls{5g}-\gls{sdr} kernels, showing \SI{0.18}-\SI{0.84}{\tera\ops} and \SI{60}-\SI{125}{\giga\ops\per\watt} in real \gls{5g} 7.X split benchmarks, meets the \gls{oran} specifications and is the first open-sourced\footnote{\url{https://github.com/pulp-platform/mempool}} programmable solution for complete physical layer processing.
}
\end{itemize}

\section{Architecture}
While it was demonstrated that shared-L1 clusters with up to \num{256} cores can be built~\cite{Mempool_2023}, their performance is still insufficient for low-latency \gls{tti} \gls{5g} pipelines. 
To meet real-life \gls{oran} requirements, TeraPool-SDR needs to "enter into uncharted territories" in terms of core count, aiming at a 4x increase over the largest cluster reported in the literature. We must aggressively leverage physical design awareness, building the cluster hierarchically to ensure high \glspl{pe}-L1 bandwidth and energy efficiency. 

\subsection{Snitch Core, Tile and Interconnection}
\label{sec:architecture}
TeraPool-SDR's \glspl{pe} are single-stage 32-bit RISC-V \textbf{\emph{Snitch}} cores~\cite{Snitch_2020} featuring an accelerator port to offload complex instructions to pipelined functional units. 
Our Snitch supports the RV32IMAXpulp- img\footnote{The Xpulpimg extension includes domain-specific instructions~\cite{Mempool_2023}, \eg \gls{mac} and load-post-increment.} \gls{isa} and multiple outstanding transactions to tolerate multi-cycle memory access latencies.
Snitch retires loads out-of-order, yet delivers in-order data to the execution units.

\begin{figure}[ht]
  \centering
  \includegraphics[width=\linewidth]{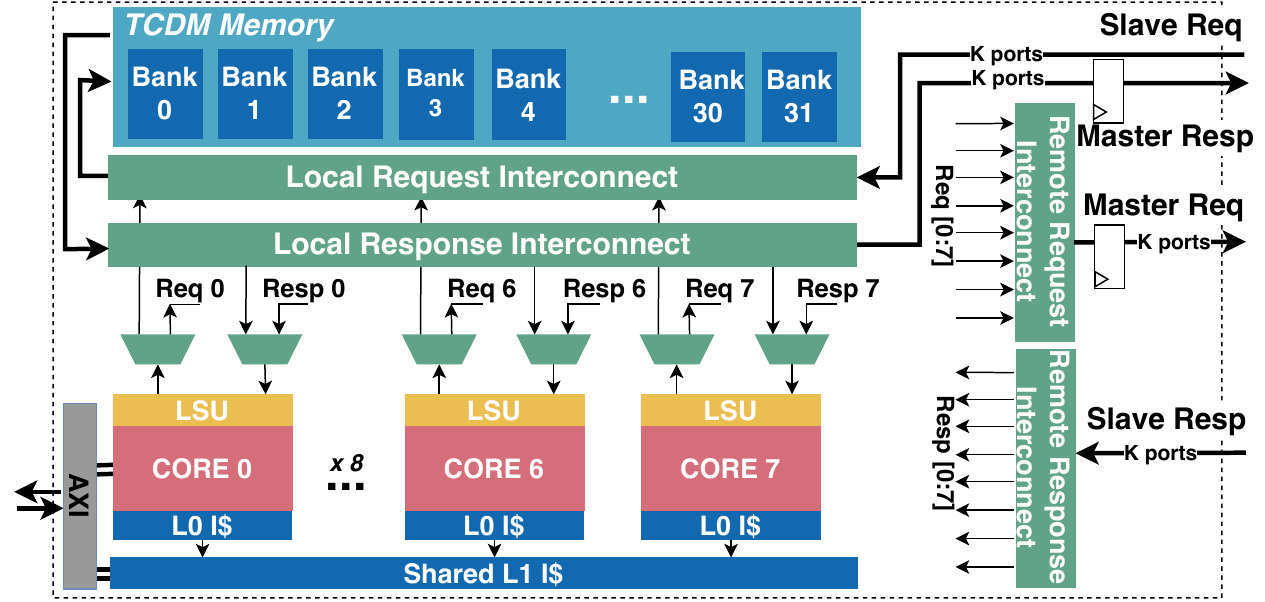}
  \caption{The TeraPool-SDR Tile architecture, the \gls{fc} crossbar interconnections protocol specified in \Cref{sec:architecture}.}
  \Description{The TeraPool-SDR Tile architecture figure shows the fully connected crossbar interconnection protocol of local and remote Tile request and response}
  \label{fig:Tile}
\end{figure}

The basic building block of TeraPool-SDR is the \textbf{\emph{Tile}}, \Cref{fig:Tile}.
It contains \num{8} Snitch \glspl{pe}, with a \num{32} instructions private L0 \gls{IDol} each, and a shared \SI{4}{KiB} two-way set-associative L1 \gls{IDol}.
The \glspl{pe} are tightly coupled to \num{32} \SI{1}{\kibi\byte} data \gls{sram} L1 banks.
The \gls{axi} master port, used for L1 \gls{IDol} refill, and \gls{dma} transfers, for data-intensive applications, is shared between all the cores.

A hierarchical topology is essential for an implementable interconnection between \num{1024} \glspl{pe} and \num{4096} memory banks.
To keep low-latency \gls{tcdm} access, we implement \gls{fc} logarithmic crossbars and arbitrators within each design level to have purely combinational routing.
Pipeline registers are optionally added at hierarchy boundaries to reduce critical paths in the physical design at the cost of increased latency.
Separate request (address, data write, and control) and response (request ID, data read, and acknowledgment) networks then handle \gls{numa} accesses.

At the Tile level, \glspl{pe} are connected to the local portion of the cluster's shared \gls{spm} with a \gls{fc} crossbar, providing single-cycle zero-load access latency.
Each Tile has a parameterizable number of remote ports to route requests to other Tiles' \gls{spm} banks via the \emph{remote request interconnect}. 
Corresponding responses are routed to the core on the \emph{remote response interconnect}. 
To access the \gls{sram} banks, incoming requests directly connect to the Tile \gls{fc} local crossbar.
The local interconnect provides a master port for each of the $\text{K}$ incoming request ports, resulting in an $(\num{8} + \text{K})\times\num{32}$ crossbar.

\subsection{Cluster Configurations}
\label{sec:configurations}
The TeraPool-SDR Cluster consists of \num{128} Tiles, including \num{1024} Snitch cores and \num{4096} \SI{1}{\kibi\byte} \gls{sram} banks of L1 \gls{spm}. Choosing the hierarchies and the placement of the interconnects within, strongly conditions the design feasibility. For example, a single upper hierarchy with \num{128} Tiles connected by a \num{128}$\times$\num{128} crossbar is clearly unfeasible. Grouping \num{16} Tiles results in $\text{K}=8$ ports per Tile (one for each Tile-Group) and delivers high inter-Tile interconnect bandwidth. However, it requires eight \num{16}$\times$\num{16} \gls{fc} crossbars within the hierarchy level, and leads to unmanageable routing congestion. Moreover, it is impossible to place the 8 Tile-Groups in a grid that results in balanced, short access paths and the \gls{eda} runtime for the Tile-Group design iteration is unmanageable. More details on physical feasibility are shown in Section~\ref{sec:physical}.

To make such a huge cluster physically feasible, we propose the following design strategies for the \terapool{} architecture:
\begin{enumerate}[leftmargin=*]
    \item  The topmost hierarchy, the Group, is replicated \num{4} times and arranged in a \num{2}$\times$\num{2} grid to both shorten and balance the diagonal access paths between the Groups.
    \item  To keep the \gls{eda} tool's runtime manageable, a Group is divided into multiple finely-tuned physical implementation hierarchies. That is, each Group consists of \num{4} SubGroups, and each SubGroup contains \num{8} Tiles.
    \item  Targeting high inter-tile bandwidth, we forward Tiles' remote Group requests directly to the Group level, where we implement \num{32}$\times$\num{32} \gls{fc} crossbars for each target Group.
\end{enumerate}
Given the interconnect required to establish local connections among Tiles in a SubGroup, we partition the local interconnection among the four SubGroups, equipping each with four \num{8}$\times$\num{8} \gls{fc} crossbars, improving physical routing feasibility. 
A Tile's remote request is routed based on the targeted hierarchy: local SubGroup, remote SubGroups, or remote Groups, resulting in $\text{K}=7$ ports: \num{1} connects locally to other Tiles in the same SubGroup via a \num{8}$\times$\num{8} \gls{fc} crossbar, \num{3} connect to Tiles in the other \num{3} SubGroups within the local Group via three \num{8}$\times$\num{8} \gls{fc} crossbars, and \num{3} connect to Tiles in the other \num{3} remote Groups via three \num{32}$\times$\num{32} crossbars.

\begin{figure*}[ht]
  \centering
  \includegraphics[width=\linewidth]{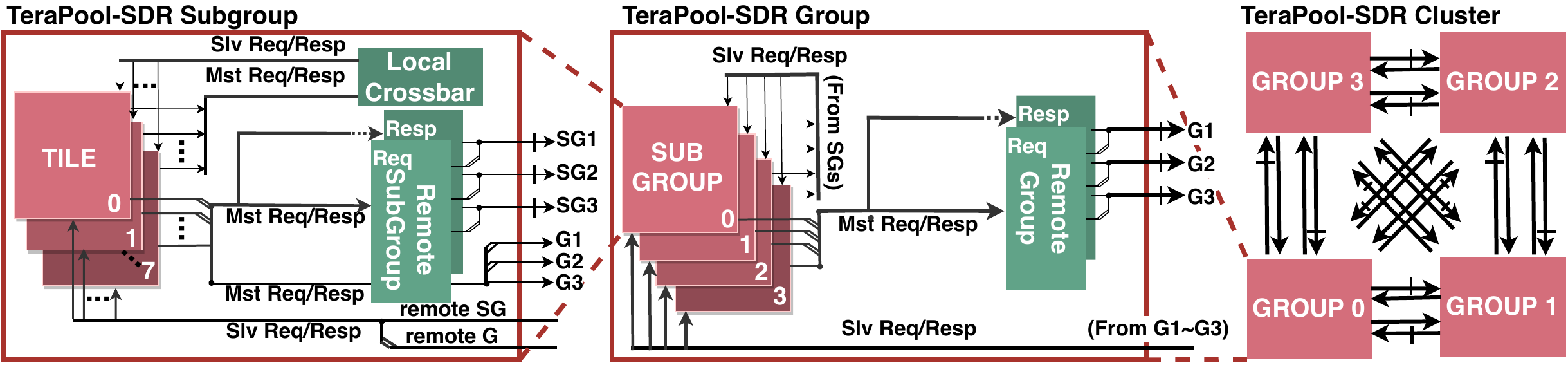}
  \caption{Bottom-up architecture of \terapool{\text{1-3-5-7}}, with the interconnection protocol specified in~\Cref{sec:architecture}.}
  \Description{Bottom-up architecture of TeraPool-SDR 1-3-5-7 configuration, with the interconnection protocol}
  \label{fig:terapool_arthitecture}
\end{figure*}

The hierarchy levels provide flexibility to place pipeline registers and break long remote access paths.
Within the Group, registers are placed at each hierarchy boundary on master ports.
The latency between Groups is a hardware-parameter that trades off the target operating frequency and latency.
We call those parametrizations \textbf{\terapool{\text{1-3-5-X}}} architecture, where the subscripts indicate the zero-load cycle latency for core access at each hierarchy level, i.e., Tile, SubGroup, Group, Cluster.
The \textbf{\terapool{\text{1-3-5-5}}} design has no extra registers between the two Groups.
A register is added on both request and response paths between Groups at the Cluster level in the \textbf{\terapool{\text{1-3-5-7}}} design, increasing the round-trip latency by \num{2} cycles.
In \textbf{\terapool{\text{1-3-5-9}}} and \textbf{\terapool{\text{1-3-5-11}}}, additional spill registers are added respectively to slave or both master-slave ports of the Group hierarchy level.
The maximum access latency increases by \num{4} and \num{6} cycles. As an example, the full architecture overview of \terapool{\text{1-3-5-7}} is shown in \Cref{fig:terapool_arthitecture}. 
These configurations create a trade-off between achievable frequency and L1 worst-case latency.
A cluster-level \gls{axi} interface is available for connection with external peripherals, chosen depending on the system design, which is not discussed in this paper.
 
\subsection{Interconnection latency-throughput trade-off}
To compare the performance of various topologies, we replace the cores with traffic generators producing memory requests following a Poisson process with a rate $\lambda$ and targeting random and uniformly distributed destination banks.
We present the latency and throughput as a function of the injected load, measured in requests per core per cycle.

\begin{figure}[htbp]
  \centering
  \begin{minipage}[ht]{0.5\linewidth}
    \centering
    \includegraphics[width=\linewidth]{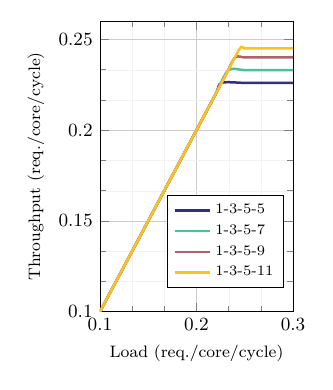}
    \label{fig:throughput}
  \end{minipage}\hfill%
  \begin{minipage}[ht]{0.5\linewidth}
    \centering
    \includegraphics[width=\linewidth]{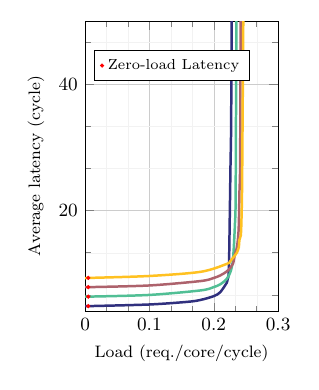}
    \label{fig:latency}
  \end{minipage}
  \caption{Throughput and average round-trip latency of TeraPool-SDR's L1 interconnect as a function of the load.}
  \Description{Throughput and average round-trip latency of TeraPool-SDR's L1 interconnect as a function of the load}
  \label{fig:latency_throughput}
\end{figure}

As shown in~\Cref{fig:latency_throughput} left, the throughput exhibits a linear-increasing trend until saturation, as the injected requests start conflicting in the interconnect to the L1 banks.
Increasing the number of pipeline registers creates a deeper interconnection that accepts more requests before generating conflicts. Therefore, the saturation load ranges \num{0.23}-\SI{0.245}{\request\per\core\per\cycle}.
\Cref{fig:latency_throughput} right shows the average round-trip latency of a memory request.
For a low injected load, the simulated latency asymptotically approaches the zero-load access latency, ranging from \num{4.9} to \num{9.3}, depending on the configuration.
As the injected load increases, the average latency grows, its asymptotic knee going from \num{10} to \num{14} (\num{40}\% variation) and corresponds to throughput saturation, at the interconnect congestion.
This analysis shows a clear trade-off between latency and throughput of different configurations. However, random loads generate an adverse access pattern with no locality within Tiles.
We analyze performance on key \gls{oran} application kernels in \cref{sec:sdr_performance}.

\section{Physical Implementation}
\label{sec:physical}
We implement TeraPool-SDR with GlobalFoundries' \SI{12}{\nano\meter} LPPLUS FinFET technology, using Synopsys' Fusion Compiler 2022.03 for synthesis, \gls{pnr}, determining the power consumption by Synopsys' PrimeTime 2022.03 under typical operating conditions (TT/\SI{0.80}{\volt}/\SI{25}{\celsius}) with switching activities obtained from post-\gls{pnr} gate-level simulations with back-annotated parasitic information. We use a bottom-up implementation flow, creating abstract design views during assembly, including only interface logic and register timing information, to reduce the \gls{eda} runtime.


\subsection{Floorplan and Feasibility}
\begin{figure}[ht]
  \centering
  \includegraphics[width=\linewidth]{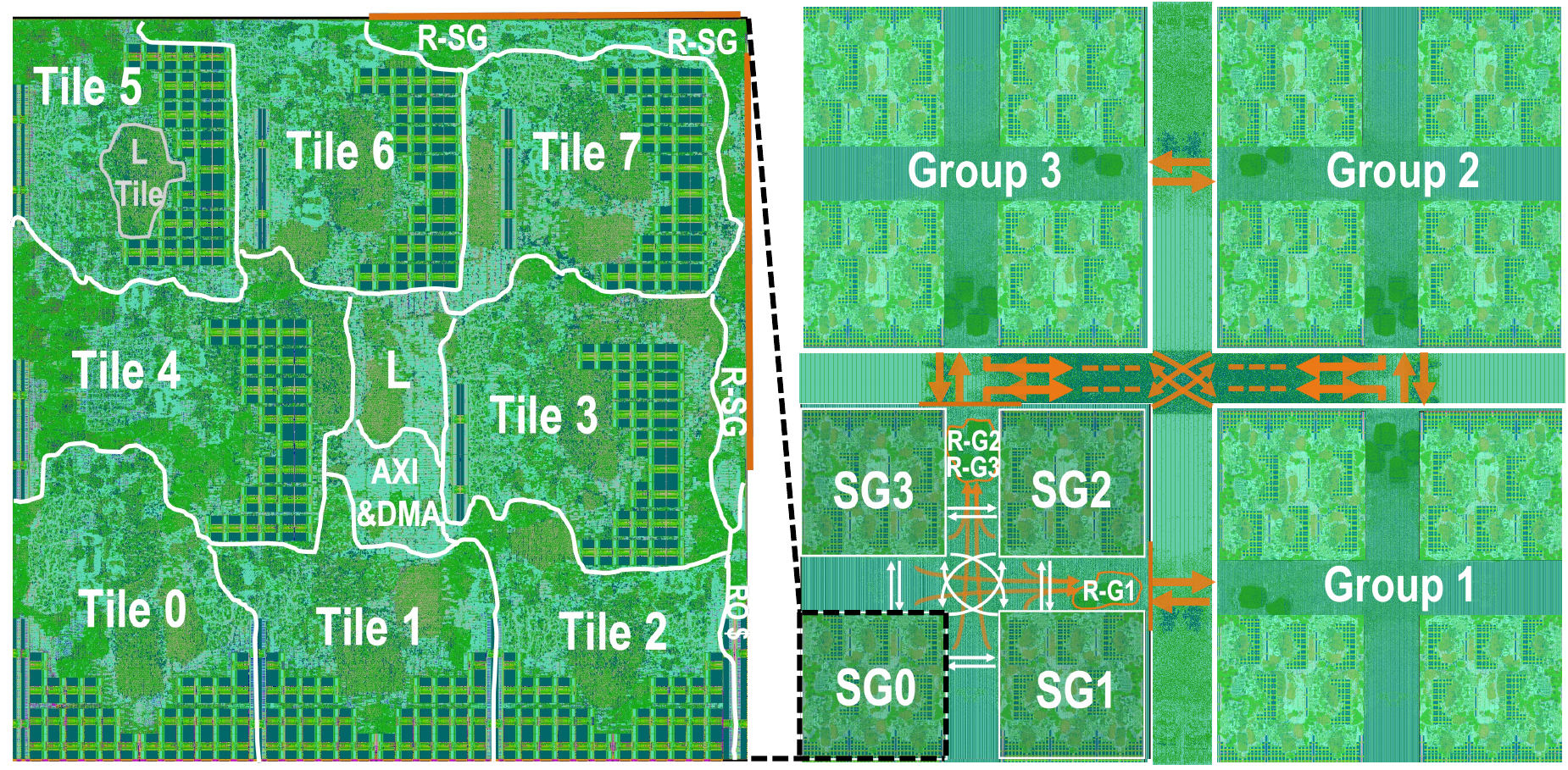}
  \caption{Placed-and-routed layout annotated view of each TeraPool-SDR hierarchical instance.}
  \Description{Placed-and-routed layout annotated view of each TeraPool-SDR hierarchical instance}
  \label{fig:terapool_layout}
\end{figure}

We show TeraPool-SDR physical design full view in ~\Cref{fig:terapool_layout}.
To fully leverage the available \gls{beol} resources and ease interconnect routing through Tiles without manual port placement, we flatten the Tile in the SubGroup, obtaining a $1.52mm \times 1.52mm$ area with a satisfactory utilization of \num{55}\%, as the first implementation hierarchy level.
The \gls{spm} macros of each Tile are grouped and arranged in a U-shape to enclose the local crossbar, minimizing overall distance and avoiding excessive stacking of macros.
SubGroup and Group blocks are arranged in a point-symmetric grid to balance the diagonal access paths, with channels in between to place and route the interconnects.
To further improve area utilization, we place interface ports behind the SubGroup blocks and shrink the channel width until \gls{beol} resources become limited.

The runtime of \gls{eda} tools serves as a fundamental indicator of both the design optimization effort and viability~\cite{Physical_hier_2017}.
As mentioned in Section~\ref{sec:configurations}, although the Cluster configuration comprising \num{8} Groups with \num{16} Tiles each delivers high inter-Tile interconnect bandwidth, it is not physically feasible.
The trail implementation shows total \gls{eda} runtime of this configuration is nearly $3.5\times$ that of \terapool{\text{1-3-5-9}}, with timing optimization accounting for more than \num{80}\% of the effort, and the routing stage is $5.5\times$ slower than the other configurations.
Despite these high design optimization efforts, the eight $16\times16$ interconnects in this Group design generate significant routing congestion and numerous metal shorts in the post-routing design phase.
Furthermore, routing detours considerably increase the length of timing paths, making it impossible to close the design with a \SI{500}{\mega\hertz} target frequency, under typical operating conditions(TT/\SI{0.80}{\volt}/\SI{25}{\celsius}).
All TeraPool-SDR configurations are physically feasible. 

\subsection{Peak-Performance \& Area}
\Cref{tab:terapool_ppa} presents the post-\gls{pnr} peak-performance \& area results for the most promising TeraPool-SDR configurations, in comparison to a smaller-scale, 256-core MemPool design in the same technology node. For fair comparison, we keep \SI{55}{\percent{}} area utilization on the base hierarchy of all the implementations.
TeraPool-SDR organizes wisely hierarchical interconnects that contribute minimal area overhead, accounting for only \num{8.5}\% of the total design area. 
TeraPool-SDR breaks the \si{\tera\ops} wall, achieving peak performance comprised between \SI{1.50}{\tera\ops} and \SI{1.89}{\tera\ops}, depending on the latency configuration.
Although a large-scale physical design leads to longer distance paths, TeraPool-SDR's low-latency scheme still keeps a small \gls{fo4} delay.
When the configured latency exceeds \num{11} cycles, the operating frequency is no longer constrained by the remote Group access path but by the Snitch core: the critical path, consisting of only \num{63} logic levels, starts at a register after the instruction cache, passes through Snitch and a request interconnection, and arrives at the clock gating of an \gls{sram} bank.
On the top hierarchy, TeraPool-SDR loses area efficiency due to the large $32\times32$ interconnection modules in each Group, requiring more routing area, and the difficult routing of feedthrough connections across Groups, caused by the blocking SubGroups placement.
These challenges impose physical constraints on the further scaling of shared-L1-memory architectures beyond the 1000-cores milepost. We nevertheless believe that exploring 3D-IC solutions will be beneficial for future research~\cite{HeterPlatform}.

\begin{table}[htb]
  \caption{Post-\gls{pnr} peak-performance \& area results.}
  \setlength{\tabcolsep}{2pt}
  \resizebox{\linewidth}{!}{%
    \begin{tabular}[h]{crcccc}
      \toprule & & \multicolumn{1}{c}{MemPool$_{256}$\textsuperscript{*}} & \multicolumn{3}{c}{TeraPool-SDR$_{1024}$\textsuperscript{*}} \\
      \cmidrule(l){3-3} \cmidrule(l){4-6} 
      
      \multicolumn{2}{r}{Hier. Access Latency [\si{\cycle}]}                            & {1-3-5}       & {1-3-5-7}     & {1-3-5-9}     & {1-3-5-11}\\\midrule
      \multicolumn{2}{r}{Area [\si{\milli\meter\squared}]}                              & \num{10.0}    & \num{68.9}    & \num{68.9}    & \num{68.9}\\
      \multicolumn{2}{r}{Area Per Core [\si{\milli\meter\squared/core}]}                & \num{0.039}   & \num{0.067}   & \num{0.067}   & \num{0.067}\\
      \multicolumn{2}{r}{Logic Gate [\si{MGE}]}                                         & \num{44}      & \num{176}     & \num{176}     & \num{176}\\
      \multicolumn{2}{r}{Logic Gate Per Core [\si{MGE/core}]}                           & \num{0.17}    & \num{0.17}    & \num{0.17}    & \num{0.17}\\\midrule
      \multicolumn{2}{r}{TCDM Spill Register in Group}                                  & \checkmark    & \checkmark    & \checkmark\checkmark & \checkmark\checkmark\checkmark\\ 
      \multicolumn{2}{r}{TCDM Spill Register in Cluster}                                & \text{-}      & \checkmark    & \checkmark    & \checkmark\\\midrule 
      \multicolumn{2}{r}{Operating Frequency \emph{(Worst)} [\si{\mega\hertz}]}         & \num{728}     & \num{530}     & \num{637}     & \num{740}\\
      \multicolumn{2}{r}{Operating Frequency \emph{(Typ.)} [\si{\mega\hertz}]}          & \num{915}     & \num{730}     & \num{880}     & \num{924}\\
      \multicolumn{2}{r}{Logic Delay [\si{FO4}]}                                        & \num{162.4}   & \num{204.5}   & \num{169.8}   & \num{161.2}\\\midrule
      \multicolumn{2}{r}{Max Throughput [\si{req/core/\cycle}]}                         & \num{0.33}    & \num{0.23}    & \num{0.24}    & \num{0.25}\\
      \multicolumn{2}{r}{Avg. Latency \emph{(Zero-load)} [\si{\cycle}]}                 & \num{4.7}     & \num{6.4}     & \num{7.9}     & \num{9.3}\\\midrule
      \multicolumn{2}{r}{Peak Performance \emph{(Typ.)} [\si{\teraops}]}                & \num{0.47}    & \num{1.50}    & \num{1.80}    & \num{1.89}\\
      \multicolumn{2}{r}{Area Efficiency [\si{\giga\ops\per\milli\meter\squared}]}      & \num{47.0}    & \num{21.8}    & \num{26.1}    & \num{27.4}\\\midrule
    \end{tabular}}
  \small\raggedright\textsuperscript{*} Subscript indicates core count; implemented in GF12LP+ technology.
  \label{tab:terapool_ppa}
\end{table}

\section{SDR Benchmarks Performance}
\label{sec:sdr_performance}

TeraPool-SDR offers a streamlined fork-join programming model, highly suitable for hardware-software co-design of \gls{sdr} workloads. Targeting optimized architecture-native kernel libraries, we adopt a low overhead C-runtime approach and distribute portions of the input data to the \glspl{pe}, according to their ID (accessible via runtime primitives). After a parallel-section cores are synchronized with a runtime barrier call. To keep utilization high and avoid contentions to shared memory resources, \glspl{pe} are constrained to fetch data from portions of the shared \glspl{spm} they can access with low latency. Key kernels for \gls{sdr} workloads are implemented: \gls{fft}, beamforming \gls{matmul}, \gls{che}, and \gls{sysinv}. 

We adopt a radix-4 decimation in frequency Cooley-Turkey \textbf{\gls{fft}}, to ease the memory accesses in local banks~\cite{PUSCH_2023_DATE}. In the $k^{th}$ stage of an N-points \gls{fft}, each core computes \num{4} butterflies, taking \num{4} inputs at a distance of $N/(4 \times 4k)$. Cores working on different \glspl{fft} are independently synchronized. 

The \textbf{\gls{matmul}} tiled implementation~\cite{PUSCH_2023_DATE}, aims to fully utilize register file in Snitch, maximizing computational intensity. The implementation reduces interconnect stalls, through a cascaded parallelization, where \glspl{pe} of the same Tile shift the fetch address start point along matrix rows, to avoid contentions for the same Group port. 

In \textbf{\gls{che}}, consisting of an element-wise matrix division, cores loop across memory rows, compute the fetch-indexes via modulo operations, and only produce the outputs residing in their local banks. 

The \textbf{\gls{sysinv}} leverages weak-scaling: as in the lower-PHY the per-subcarrier \gls{mimo} has less than \num{32} transceivers we assign an independent small squared \gls{sysinv} problem to each core and then synchronize. We use \gls{choldec} to invert the linear system, and store the intermediate lower-triangular decomposition in the local memory of the core working on it.

The large capacity of TeraPool-SDR's L1 allows to set up functional pipelines without resorting to L2 accesses to solve the data dependencies between the different steps of lower-PHY processing (\eg in a typical \gls{ofdm} workload the data of \gls{fft} for up to 64 antennas can be passed to the \gls{matmul} beamforming stage with no L2 memory transfers).

\begin{table}[htb]
  \caption{Performance metrics of \gls{sdr} key-kernels.}
  \setlength{\tabcolsep}{2pt}
  \resizebox{\linewidth}{!}{%
    \begin{tabular}[h]{crccc}
      \toprule & & \multicolumn{1}{c}{\tp{\text{1-3-5-7}}} & \multicolumn{1}{c}{\tp{\text{1-3-5-9}}} & \multicolumn{1}{c}{\tp{\text{1-3-5-11}}} \\
      \cmidrule(l){3-3} \cmidrule(l){4-4} \cmidrule(l){5-5} 
      \multirow{4}{*}{FFT}     & {Max Size in L1}  & {\numproduct{64x4096}}   & {\numproduct{64x4096}}   & {\numproduct{64x4096}}   \\
                               & {Total Cycles}    & {\num{31740}}            & {\num{32012}}            & {\num{32272}}            \\
                               & {IPC}             & {\num{0.75}}             & {\num{0.75}}             & {\num{0.74}}             \\ \midrule
      \multirow{4}{*}{MatMul}  & {Max Size in L1}  & {\numproduct{512x512}}   & {\numproduct{512x512}}   & {\numproduct{512x512}}   \\
                               & {Total Cycles}    & {\num{296995}}           & {\num{298239}}           & {\num{301113}}           \\
                               & {IPC}             & {\num{0.70}}             & {\num{0.69}}             & {\num{0.69}}             \\ \midrule
      \multirow{4}{*}{CHE}     & {Max Size in L1}  & {4096 \numproduct{32x4}} & {4096 \numproduct{32x4}} & {4096 \numproduct{32x4}} \\
                               & {Total Cycles}    & {\num{16851}}            & {\num{17221}}            & {\num{17601}}            \\
                               & {IPC}             & {\num{0.66}}             & {\num{0.64}}             & {\num{0.63}}             \\ \midrule
      \multirow{4}{*}{SysInv}  & {Max Size in L1}  & {65536 \numproduct{4x4}} & {65536 \numproduct{4x4}} & {65536 \numproduct{4x4}} \\
                               & {Total Cycles}    & {\num{30539}}            & {\num{31254}}            & {\num{31870}}            \\
                               & {IPC}             & {\num{0.61}}             & {\num{0.59}}             & {\num{0.58}}             \\ \midrule
    \end{tabular}}
  \small\raggedright\textsuperscript{*}Each core supports 8 outstanding transactions in the each given system.
  \label{tab:software}
\end{table}

\begin{figure}[ht]
  \centering
  \includegraphics[width=\linewidth]{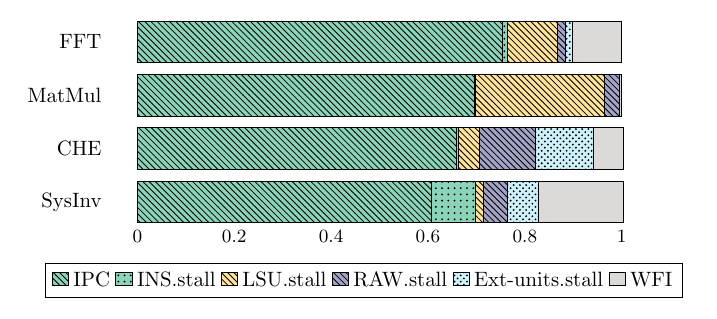}
  \caption{Fraction of instructions and stalls over the total cycles for the kernels execution in \terapool{\text{1-3-5-7}}.}
  \Description{Fraction of instructions and stalls over the total cycles for the kernels execution in TeraPool-SDR 1-3-5-7}
  \label{fig:software_ipc}
\end{figure}

\Cref{tab:software} demonstrates that TeraPool-SDR's large shared-L1 memory enables processing of the entire data frame in a typical \gls{5g} use case without splitting, achieving an \gls{ipc} of up to 0.75 for key \gls{sdr} kernels.
TeraPool-SDR's remote Group access latency varies from 7 to 11 cycles, but the \gls{ipc} loss across different configurations stays below \num{3}\percent{}, proving an effective latency-hiding in software and hardware.
\Cref{fig:software_ipc} displays the detailed \gls{ipc} and stall fractions for the execution of \gls{sdr} kernels on TeraPool-SDR. 
The parallelization scheme, minimizes \gls{lsu} stalls by eliminating access to remote memory banks for \gls{fft}, \gls{che}, and \gls{sysinv}.
The \gls{raw} and external-unit stalls in \gls{che} and \gls{sysinv} result from multi-cycle instructions (mainly divisions) offloaded to pipelined functional units during intensive computing loops.
Being a control-heavy task, \gls{sysinv} experiences more \gls{IDol} misses and synchronization overhead.
For the \gls{matmul}, although target data structures distribute over all the banks of the shared \gls{spm}, causing unavoidable \gls{lsu} stalls, the \gls{ipc} on the TeraPool-SDR Cluster still achieves \num{0.7}, highlighting that TeraPool-SDR delivers a significant fraction of the peak workload even for kernels with \gls{pe}-to-L1 traffic patterns that are not highly local, with non-negligible contention.

\begin{figure}[ht]
  \centering
  \includegraphics[width=\linewidth]{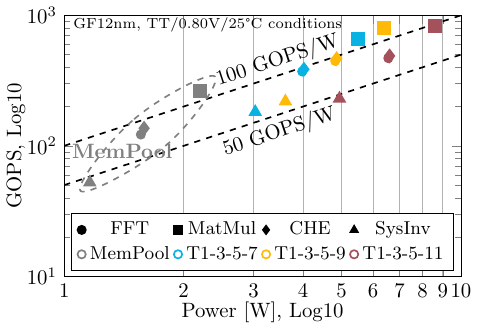}
  \caption{Energy Efficiency for SDR workloads.}
  \label{fig:terapool_energy_eff}
  \Description{Energy Efficiency for SDR workloads, shows that TeraPool-SDR achieves comparable or higher energy efficiency than a 4 times smaller-scale cluster on all the key SDR kernels}
\end{figure}

\Cref{fig:terapool_energy_eff} shows that TeraPool-SDR achieves comparable or higher energy efficiency than a \num{4}$\times$ smaller-scale cluster on all the key \gls{sdr} kernels thanks to its low-power interconnect. \terapool{\text{1-3-5-7}} is the lowest power consumption design: a memory request crossing the cluster takes \SI{13.5}{pJ}, only 0.5$\times$ more than a local request. The performance and energy efficiency of this design are however dominated by the other two.
\terapool{\text{1-3-5-11}} is Pareto-optimal in operating frequency and performance, at the cost of a higher power consumption (\SI{6.5}{\watt} for \gls{fft}, \SI{8.8}{\watt} for \gls{matmul}, \SI{6.6}{\watt} for \gls{che} and \SI{4.9}{\watt} for \gls{sysinv} kernels). With minimal performance losses, \terapool{\text{1-3-5-9}} is optimal in energy efficiency for the selected \gls{sdr} workloads. Being the most compute-intensive task, \gls{matmul} achieves a peak of \SI{125}{\giga\ops\per\watt} and consumes less than \SI{6.4}{\watt}, in compliance with the low power consumption standards of base-station hardware.

\section{Conclusion}
TeraPool-SDR is a physically feasible, many-core cluster of \num{1024} Snitch RISC-V cores sharing \SI{4}{MiB} of L1 \gls{spm}, through hierarchical low-latency \gls{numa} \gls{tcdm} interconnections (less than \num{5} cycles within local Group, \num{7}-\num{11} to remote Group, depending on the target frequency). Completing \gls{pnr} by using GlobalFoundries' 12LPPLUS FinFET technology, TeraPool-SDR occupies a die area of \SI{68.9}{\mm^2} and operates at up to \SI{924}{MHz} (\num{63} gate delays) in typical operating conditions (TT/\SI{0.80}{\volt}/\SI{25}{\celsius}) with a theoretical peak performance of \SI{1.89}{\tera\ops}.

TeraPool-SDR's large \gls{pe} count exceeds the state-of-the-art by \num{4}$\times$ and achieves high performance. Depending on the target operating frequencies, TeraPool-SDR achieves \SI{0.18}-\SI{0.84}{\tera\ops} and \SI{60}-\SI{125}{\giga\ops\per\watt} on the benchmarks of \gls{5g} 7.X splits targeted for acceleration in base-station hardware by industry-leading solutions. Thanks to its low-latency interconnect with a large shared-L1, TeraPool-SDR minimizes the data splitting, transfer, and synchronization overheads of memory-intensive \gls{sdr} applications.

\begin{acks}
This work is funded in part by the COREnext project supported by the EU Horizon Europe research and innovation program under grant agreement No. \num{101092598}.
\end{acks}

\bibliographystyle{ACM-Reference-Format}
\bibliography{terapool}

\end{document}